\documentclass[10pt]{article}
\usepackage[margin=1in]{geometry}
\usepackage{amsmath}
\usepackage{abstract}
\usepackage{amsfonts,color}
\usepackage{amssymb}
\usepackage{graphicx}
\usepackage{hyperref}
\usepackage{bigints}
\usepackage{mathrsfs}
\begin{document}
\title{\bf{Acoustic Analogue of Gravitational Wave}}
\author
{Satadal Datta\\
{\small\textit{Harish-Chandra research Institute, HBNI Chhatnag Road, Jhunsi, Allahabad-211019, India}}\\
\date{}
\small Email: satadaldatta1@gmail.com\\\small satadaldatta@hri.res.in}
\maketitle
\begin{abstract}
We explore nonlinear perturbations in different static fluid systems. We find that the equations, corresponding to the perturbation of the integrals of motion, i.e; Bernoulli's constant and the mass flow rate, satisfy massless scalar field equation in a time dependent acoustic metric. When one is interested up to the second order behaviour of the perturbations, the emergent time dependent acoustic metric of the system, derived from the massless scalar field equations of the perturbations of the integrals of motion, has some astounding similarities with the metric describing gravitational wave in Minkowski spacetime. 
\end{abstract}
\section{Introduction}
Detection of gravitational wave \cite{a} is not only one of the greatest achievement of this century in Physics but also it's an another confirmation of the general theory of relativity \cite{b}. Unruh's pioneering work \cite{c} shows that the linear perturbation of velocity potential in an inviscid irrotational fluid medium behaves like a massless scalar field propagating in a curved spacetime. Several features of classical black hole can be mimicked in different fluid systems \cite{d}. There are also some works which show that instead of linear perturbation of velocity potential, one can work with linear perturbation of the integrals of motion of the fluid equations for irrotational inviscid medium, i.e; Bernoulli's constant and mass flow rate \cite{e}-\cite{j}. However it is evident that to mimic the Minkowski spacetime, the fluid medium has to be static, i.e; the background velocity of the fluid medium has to be zero everywhere and in such a medium, the propagation of linear perturbation of velocity potential would satisfy massless scalar field equation in the acoustic analogue of Minkowski spacetime.\\ \cite{k}-\cite{m} study nonlinear perturbations in a moving fluid medium, more specifically, in an accreting medium, i.e; there is a position dependence of velocity. We consider apparently simpler fluid systems where there is no background velocity in the medium. Hence the emergent spacetime is analogous to Minkowski spacetime when one works with linear perturbations. Here we have shown that the emergent spacetime corresponding to the massless scalar field equation satisfied by the perturbation of the integrals of motion gives time dependent spacetime metric in general and when the expressions of the perturbations are expanded up to the second order, the emergent spacetime metric is very similar to spacetime metric describing gravitational wave propagating in Minkowski spacetime. There are some differences also; here the acoustic analogue of gravitational wave, propagating with the speed of sound in the medium, is not transverse in kind rather is of longitudinal type. Our work explores the longitudinal wave nature of the acoustic metric in different fluid systems.        
\section{Nonlinear acoustics}
We first consider the simplest possible system, a medium of uniform density, $\rho_0$ and of uniform pressure, $p_0$ i.e; ${\bf{\nabla}}\rho_0\sim 0,~{\bf{\nabla}} p_0\sim 0$. The effect of any external field is assumed to be negligible. The fluid equations for inviscid flow read as,
\begin{equation}
\frac{\partial\rho}{\partial t}+{\bf{\nabla}.(\rho{\bf{v}})}=0
\end{equation}
\begin{equation}
\frac{\partial \bf{v}}{\partial t}+{\bf{v}}.{\bf{\nabla}}{\bf{v}}=-\frac{{\bf{\nabla}}p}{\rho}
\end{equation}
We introduce perturbations, not necessarily linear, as 
\begin{align*}
& p({\bf{x}}, t)=p_0+p'({\bf{x}}, t)\\
&\rho({\bf{x}}, t)=\rho_0+\rho'({\bf{x}}, t)\\
&{\bf{v}}({\bf{x}}, t)={\bf{v}}'({\bf{x}}, t)\\
\end{align*}
The motion of the fluid is assumed to be irrotational, i.e; ${\bf{\nabla}}\times{\bf{v}}={\bf{\nabla}}\times{\bf{v'}}=0$. 
Hence
\begin{equation}
\frac{\partial {\bf{v}}}{\partial t}+{{\bf{\nabla}}\left(\frac{1}{2}{\bf{v}}^2+\int\frac{dp}{\rho}\right)}=0
\end{equation}
For steady state problem, the conserved quantity derived from the momentum equation is Bernoulli's constant, $\zeta$ and $\zeta=\left(\frac{1}{2}{\bf{v}}^2+\int\frac{dp}{\rho}\right)$ in general. Equation (3) implies
\begin{equation}
\frac{\partial {\bf{v'}}}{\partial t}+{{\bf{\nabla}}\zeta'}=0
\end{equation}
One can write $\zeta$ as
\begin{equation}
\zeta({\bf{x}},t)=\zeta_0+\zeta'({\bf{x}},t)
\end{equation}
where $\zeta_0$ corresponds to the background value of the Bernoulli's constant, which is a constant number and $\zeta'({\bf{x}},t)$ is the nonlinear fluctuation around this value. 
From the expression, 
\begin{equation}
\partial_t\zeta'={\bf{v'}}.\partial_t{\bf{v'}}+\frac{c_{s}^2}{\rho}\partial_t\rho'
\end{equation}
where, $c_{s}^2=\frac{dp}{d\rho}$ from definition.
Using the continuity equation, equation (1); and Euler momentum equation, equation (2);
\begin{equation}
\partial_\mu (f^{\mu\nu}({\bf{x}},t)\partial_\nu )\zeta'({\bf{x}},t)=0
\end{equation}
where 
\begin{equation}
f^{\mu\nu}({\bf{x}},t)\equiv\frac{\rho}{c_{s}^2}\begin{bmatrix}
-1 & \vdots & -v'^{j} \\
\cdots&\cdots&\cdots\cdots \\
-v'^{j}&\vdots & c_{s}^{2}\delta^{ij}-v'^{i}v'^{j}
\end{bmatrix}
\end{equation}
indices $i,~j$ run from $1$ to $3$ and the Greek indices run from $0$ to $3$. Now one can find the time dependent acoustic metric by comparing the above equation with the massless scalar field equation as follows.
\begin{equation}
f^{\mu\nu}=\sqrt{-g}g^{\mu\nu}
\end{equation}
where $g$, the determinant of the metric $g_{\mu\nu}$, is equal to $-\frac{\rho^4}{c_s^2}$. We find 
\begin{equation}
g_{\mu\nu}({\bf{x}},t)\equiv\frac{\rho}{c_s}\begin{bmatrix}
 -(c_{s}^{2}-v'^{2}) & \vdots & -v'^{i} \\
\cdots&\cdots&\cdots\cdots \\
-v'^{i}&\vdots &\delta_{ij}
\end{bmatrix}
\end{equation} 
Hence the perturbation of Bernoulli's constant satisfies massless scalar field equation in a time dependent spacetime \footnote{Instead of working with $\zeta'$, one can work with $\zeta$ because $\zeta_0$ is just a constant number.}.\\
$\zeta'({\bf{x}},t),~f^{\mu\nu}({\bf{x}},t)$ can be written as
\begin{align*}
& \zeta'({\bf{x}},t)=\zeta'_{(1)}({\bf{x}},t)+\zeta'_{(2)}({\bf{x}},t)+...\\
& f^{\mu\nu}({\bf{x}},t)=f^{\mu\nu}_{(0)}+f^{\mu\nu}_{(1)}({\bf{x}},t)+...
\end{align*}
where $\zeta'_{(n)},~f^{\mu\nu}_{(n)}$ s, $n$ being natural number, are the $n$th order term in smallness in the expansion of $\zeta'({\bf{x}},t)$ and $f^{\mu\nu}({\bf{x}},t)$ respectively. In the expansion of $f^{\mu\nu}({\bf{x}},t)$,

 $f^{\mu\nu}_{(0)}$ is the zeroth order term depending on the background values of density and sound speed in the medium. Writing the equation (7) upto second order,
\begin{equation}
\partial_\mu (f^{\mu\nu}_{(0)}+f^{\mu\nu}_{(1)}({\bf{x}},t))\partial_\nu\left(\zeta'_{(1)}({\bf{x}},t)+\zeta'_{(2)}({\bf{x}},t)\right)+O(3)+...=0
\end{equation} 
In the first order of smallness, $\zeta'_{(1)}$ satisfies the usual known equation, $\partial_\mu (f^{\mu\nu}_{(0)}({\bf{x}},t)\partial_\nu )\zeta'_{(1)}({\bf{x}},t)=0$; from this equation one can find the acoustic metric analogous to Minkowski metric in flat spacetime. Now we neglect $O(3)$ and higher order terms. Rewriting equation (11) upto second order 
\begin{equation}
\partial_\mu (\tilde{f}^{\mu\nu}({\bf{x}},t)\partial_\nu )\tilde{\zeta'}({\bf{x}},t)=0
\end{equation}
where $\tilde{f}^{\mu\nu}({\bf{x}},t)$ is $f^{\mu\nu}({\bf{x}},t)$ expanded upto first order of smallness and $\tilde{\zeta'}({\bf{x}},t)$ is $\zeta'({\bf{x}},t)$ expanded upto second order in smallness \footnote{The last term is a $O(3)$ term, here we are treating $\partial_\mu (\tilde{f}^{\mu\nu}\partial_\nu)$ as an operator acting on the perturbation $\zeta'$ and so to retain the form of the equation in operator form we are taking one of the $O(3)$ term using weakly nonlinear limit; one can of course equate each term corresponding to several degrees of smallness to zero order by order but here our approach is different, we neglect other $O(3)$ terms anyway by equating the expression approximately to zero in operator form.}. 
Similarly perturbation of other fluid quantities can be expressed as below.
\begin{align*}
& \rho({\bf{x}},t)=\rho_0+\rho'_{(1)}({\bf{x}},t)+...\\
& p({\bf{x}},t)=p_0+p'_{(1)}({\bf{x}},t)+...\\
& c_s({\bf{x}},t)=c_{s0}+c_{s (1)}({\bf{x}},t)+...\\
& c_{s}^2({\bf{x}},t)= c_{s0}^2+c_{s (1)}^2({\bf{x}},t)+...\\
& {\bf{v}}'({\bf{x}}, t)={\bf{v}}'_{(1)}({\bf{x}}, t)+{\bf{v}}'_{(2)}({\bf{x}}, t)+...\\
\end{align*}
where $c_{s0}$ is the sound speed, i.e; the speed of propagation of linear perturbations. 
To study second order behaviour, we expand $f^{\mu\nu}({\bf{x}},t)$ upto first order, hence $g_{\mu\nu}({\bf{x}}, t)$ has to be expanded upto linear order as
\begin{equation}
g_{\mu\nu}({\bf{x}}, t)=g_{(0)\mu\nu}+g_{(1)\mu\nu}({\bf{x}}, t)
\end{equation} 
Assuming isentropic perturbations in this isothermal medium \footnote{the medium is isothermal in the absence of any perturbation because the medium is taken to be uniform in density and pressure, hence from equation of state of ideal gas, the medium has to be isothermal.}, we use barotropic equation for fluid.
\begin{equation}
p=K\rho^{\gamma}
\end{equation}
where $K$ is a constant number and $\gamma$ is the specific heat ratio. Therefore when one works with perturbations upto first order in smallness, the speed of propagation of linear perturbation, i.e; the adiabatic sound speed, $c_{s0}$, is given by \cite{n}
\begin{equation}
c_{s0}^2=\frac{p'_{(1)}({\bf{x}}, t)}{\rho'_{(1)}({\bf{x}}, t)}=\frac{\gamma p_0}{\rho_0}
\end{equation}
One can work with isothermal perturbation also, for sound wave propagating in the air medium, the adiabatic approximation works much better than the isothermal one \cite{n}.\\
Using equation (14), we find that 
\begin{align*}
& \frac{c_{s (1)}({\bf{x}},t)}{c_{s0}}=\frac{(\gamma-1)}{2}\frac{\rho'_{(1)}({\bf{x}},t)}{\rho_0}\\
& \frac{c_{s (1)}^2({\bf{x}},t))}{c_{s0}^2}=(\gamma-1)\frac{\rho'_{(1)}({\bf{x}},t)}{\rho_0}
\end{align*}  
Using the above expressions in the matrix of the equation (10), we get
\begin{equation}
g_{\mu\nu}({\bf{x}}, t)=g_{(0)\mu\nu}+g_{(1)\mu\nu}({\bf{x}}, t)\equiv\frac{\rho_0}{c_{s0}}\begin{bmatrix}
 -c_{s0}^{2}\left(1+\frac{(\gamma+1)}{2}\frac{\rho'_{(1)}({\bf{x}},t)}{\rho_0}\right) & \vdots & -v'^{i}_{(1)}({\bf{x}}, t) \\
\cdots&\cdots&\cdots\cdots \\
-v'^{i}_{(1)}({\bf{x}}, t)&\vdots &\delta_{ij}\left(1+\frac{(3-\gamma)}{2}\frac{\rho'_{(1)}({\bf{x}},t)}{\rho_0}\right)
\end{bmatrix}
\end{equation}  
As $\frac{\rho_0}{c_{s0}}$ is just a constant number in front of the above matrix. We work with a better looking matrix, defined by
\begin{equation}
\tilde{g}_{\mu\nu}({\bf{x}}, t)=\frac{c_{s0}}{\rho_0}g_{\mu\nu}({\bf{x}}, t)=(\eta_A)_{\mu\nu}+h_{\mu\nu}({\bf{x}}, t)
\end{equation}
where $(\eta_A)_{\mu\nu}$, the acoustic analogue of Minkowski metric, is: $({\rm diag}[-c_{s0}^2,+1,+1,+1])_{\mu\nu}$; the convention in \cite{o} is used. $h_{\mu\nu}$ is the linear perturbation term of the acoustic metric.\\
Now we examine the behviour of the $h_{\mu\nu}({\bf{x}}, t)$. We have from equation (1) and equation (2),
\begin{equation}
\partial_t^2\rho'_{(1)}({\bf{x}}, t)=c_{s0}^2{\bf{\nabla}}^2\rho'_{(1)}({\bf{x}}, t)
\end{equation}
and
\begin{equation}
\partial_t^2v'^{i}_{(1)}({\bf{x}}, t)=c_{s0}^2\partial_i\left(\partial_jv'^{j}_{(1)}({\bf{x}}, t)\right)
\end{equation}
We assume in our coordinate system, the $z$ component of the linear perturbation of velocity is the only non-zero component, i.e; we are studying nonlinear sound wave propagating parallel to $z$ axis. Hence
\begin{align}
& v'^{1,2}_{(1)}=0\\
& v'^3_{(1)}=v'^3_{(1)}(z,t)\\
& \rho'_{(1)}=\rho'_{(1)}(z,t)
\end{align} 
This assumption is compatible with the irrotationality condition. This is very similar to working in the harmonic coordinate system \cite{p}, i.e; choosing Einstein gauge \cite{q}, in the case of studying real gravitational wave propagating parallel to $z$ axis. Therefore, in this coordinate system, using equation (18) and equation (19), we get
\begin{equation}
\square_A h_{\mu\nu}(z,t)=0
\end{equation}
where $\square_A$ is the acoustic analogue of d'Alembertian wave operator, given by
\begin{align*}
\square_A=-\frac{1}{c_{s0}^2}\frac{\partial^2}{\partial t^2}+\bf{\nabla}^2
\end{align*}  
Hence $h_{\mu\nu}(z,t)$ represents acoustic analogue of gravitational wave propagating parallel to $z$ axis. In the case of sound wave propagating uniformly in all the directions, we would have chosen to work in the spherical polar coordinate system as harmonic coordinate system. We are doing two things simultaneously, one is that we are giving the wave vector of sound a certain direction (sound wave propagating uniformly in all directions or sound wave propagating along a particular direction etc) and we are considering suitable coordinate system to describe it. As a result the d'Alembertian operator of equation (23) happens to be $1+1$ dimensional. One can also work in other coordinate systems to get equation (23), for example in a coordinate system such that
$v'^{1,2}_{(1)}$ are constant numbers. This is similar to gauge freedom which we have in the case of real gravitational wave. \\
Therefore, $h_{\mu\nu}(z,t)$ is given by
\begin{equation}
h_{\mu\nu}(z, t)\equiv\begin{bmatrix}
 -c_{s0}^{2}\frac{(\gamma+1)}{2}\frac{\rho'_{(1)}(z,t)}{\rho_0} & 0 & 0 & -v'^3_{(1)}(z,t)\\
0 & \frac{(3-\gamma)}{2}\frac{\rho'_{(1)}(z,t)}{\rho_0} & 0 & 0\\
0 & 0 &  \frac{(3-\gamma)}{2}\frac{\rho'_{(1)}(z,t)}{\rho_0} & 0\\
-v'^3_{(1)}(z,t) & 0 & 0 &\frac{(3-\gamma)}{2}\frac{\rho'_{(1)}(z,t)}{\rho_0}
\end{bmatrix}
\end{equation}
Instead of Bernoulli's constant, one can start with studying perturbation of mass flow rate (Appendix Section), the conserved quantity derived from the continuity equation when the motion of the fluid medium is assumed to be steady. In that case, in the very beginning, one has to assume the direction of the sound wave and a suitable coordinate system to describe it, one would find similarly analogue of gravitational wave in the Minkowski spacetime.\\
Instead of adiabatic sound in the isothermal medium, one could start with isothermal sound in such a medium. Sound is approximately adiabatic in nature in air medium \cite{n}. In the case of isothermal sound, the expressions in the previous equations would be the same except one has to put $\gamma=1$.\\
Just like gravitational wave propagating in the $z$ direction, the acoustic analogue of gravitational wave has also two independent nontrivial components, $h_{00}(z,t)$ (proportional to the linear perturbation of density of the medium) and $ h_{03}(z,t)$ (proportional to the linear perturbation of velocity along $z$ axis in the medium). Other nontrivial components are derivable from these two independent nonzero components as follows;
\begin{align}
& h_{11}=h_{22}=h_{33}=-\frac{(3-\gamma)}{c_{s0}^2(\gamma+1)}h_{00}\\
& h_{30}=h_{03}
\end{align}
Equation (26) is compatible with the symmetric properties of the acoustic metric.
For sound wave propagating along $x$ axis, $h_{\mu\nu}=h_{\mu\nu}(x,t)$, the nontrivial diagonal quantities will be similar in looking as before, $h_{01}=h_{10}=v'^{1}_{(1)}(x,t)$ will be nontrivial component instead of $h_{03}$. This same conclusion can be drawn similarly by applying rotation operator over $h_{\mu\nu}$. As acoustic metric is invariant under rotation, we have
\begin{equation}
h_{\mu\nu}'=R_\mu^{~\rho} R_\nu^{~\sigma} h_{\rho\sigma}
\end{equation}
where $R_\mu^{~\rho}$ denotes rotation operator and $h'_{\mu\nu}$ denotes the linear perturbation term in the acoustic metric after rotation; writing the matrix corresponding to $h_{\mu\nu}$ as $\hat{h}$ and rotation operator as $\hat{R}$, we get
\begin{equation}
\hat{h}'=\hat{R}^T\hat{h}\hat{R}
\end{equation}
where $\hat{R}^T\hat{R}=\hat{R}\hat{R}^T=\mathbb{I}$, $\hat{R}^T$ is the transpose of $\hat{R}$ and $\mathbb{I}$ is the $4\times 4$ identity matrix. The sound wave propagating along $z$ direction in one coordinate system, has direction along $x$ axis in another coordinate system; the second coordinate system can be obtained from the first one by a rotation of $-90^\circ$ about $y$ axis. For a rotation of angle $\theta$ about $y$ axis over the matrix of equation (24), we get
\begin{equation}
\hat{h}'=\hat{R}_y^T\hat{h}\hat{R}_y=\begin{bmatrix}
 -c_{s0}^{2}\frac{(\gamma+1)}{2}\frac{\rho'_{(1)}}{\rho_0} & v'^3_{(1)}sin\theta & 0 & -v'^3_{(1)}cos\theta\\
v'^3_{(1)}sin\theta & \frac{(3-\gamma)}{2}\frac{\rho'_{(1)}}{\rho_0} & 0 & 0\\
0 & 0 &  \frac{(3-\gamma)}{2}\frac{\rho'_{(1)}}{\rho_0} & 0\\
-v'^3_{(1)}cos\theta & 0 & 0 &\frac{(3-\gamma)}{2}\frac{\rho'_{(1)}}{\rho_0}
\end{bmatrix}
\end{equation} 
Where $\hat{R_y}$ denotes rotation operator corresponding to rotation about $y$ axis. $\theta=-90^\circ$ in the above matrix equation gives desired $\hat{h}'$. Therefore by rotation about $y$ axis, the diagonal elements in the $\hat{h}$ matrix remain same. Wherefore, the diagonal entries in the $\hat{h}$ matrix are proportional to the linear perturbation in density, the diagonal elements do not change under rotation of any kind, these quantities are scalar quantities under rotation. The nontrivial off-diagonal quantities are proportional to the linear perturbation in velocity, hence they transform under rotation. \\
However, rotation about $z$ axis on $\hat{h}$ do not have any effect. This is not the case for real gravitational wave propagating along $z$ axis, the physically significant terms in the case of real gravitational wave have helicity  $\pm 2$ \cite{p}.\\
The solution of equation (23), has the general form
\begin{equation}
h_{\mu\nu}(z,t)=h_{\mu\nu}(z\pm c_s t)
\end{equation}
The speed of analogue gravitational wave is the sound speed; $'+'$ sign implies sound wave propagating along negative $z$ axis and $'-'$ sign implies sound wave propagating along positive $z$ axis. 
For a plane wave propagating along $+z$ axis, we write
\begin{equation}
h_{\mu\nu}(z,t)=e_{\mu\nu}~exp\left(i(kz-\omega t)\right)
\end{equation}
$e_{\mu\nu}$ is the amplitude of the wave. $\omega$ and $k$ are the angular frequency and the wave vector of the plane wave. Using equation (23), we find dispersion relation as
\begin{equation}
\omega=c_s k
\end{equation}
This linear dispersion relation of acoustic gravitational wave is similar to that for real gravitational wave.\\
Hence when one extends the analysis of introducing perturbations to second order, the emergent metric gets to have some striking similarities with the real gravitational wave and there are some differences as well. 
\subsection{Implications in nonlinear acoustics}
The above formalism of nonlinear perturbations can be used to understand some nonlinear phenomena. We have assumed the fluid to be inviscid even in the presence of perturbations and also there is no heat conduction or convection happening in the system. Hence  we can describe nonlinear acoustics in lossless fluids \cite{r}. Let's consider, irrotational flow along $x$ axis. The velocity, $v^1=v'^1=-\psi_x$, where $\psi$ is the velocity potential and $\psi_x$ is the partial derivative with respect to $x$.  From equation (4)
\begin{equation}
\psi_t=\zeta'
\end{equation} 
Using expression of $\zeta$, equation (6), continuity equation and the above expression, we get
\begin{equation}
\psi_{tt}-2\psi_{xt}\psi_x+(\psi_x^2)\psi_{xx}=c_{s}^2\psi_{xx}
\end{equation} 
This is the nonlinear acoustic wave equation in lossless scenario in terms of velocity potential \cite{r}-\cite{u}. Using barotropic equation (equation 14) and expression of $\zeta$, we have
\begin{eqnarray}
\zeta
=\frac{1}{2}\psi_x^2+\frac{c_s^2}{(\gamma-1)}\nonumber\\
=\zeta_0+\zeta'(x,t)\nonumber\\
=\frac{c_{s0}^2}{(\gamma-1)}+\psi_t
\end{eqnarray}
Therefore, using expression (35) in equation (34), we find
\begin{equation}
\psi_{tt}-c_{s0}^2\psi_{xx}=2\psi_{xt}\psi_x+(\gamma-1)\psi_{xx}\psi_t-\frac{(\gamma+1)}{2}\psi_x^2\psi_{xx}
\end{equation}
Furthermore, using expression (35), one can write density as a function of partial derivatives of velocity potential \cite{t},
\begin{equation}
\rho=\rho_0\left(1+\frac{(\gamma-1)}{c_{s0}^2}\left(\psi_t-\frac{\psi_x^2}{2}\right)\right)^{\frac{1}{\gamma-1}}
\end{equation} 
Neglecting smallness of cubic order, one can derive some interesting lossless wave equations in weakly nonlinear limit \cite{v},
\begin{equation}
\psi_{tt}-c_{s0}^2\psi_{xx}=2\psi_{xt}\psi_x+(\gamma-1)\psi_{xx}\psi_t
\end{equation}
Approximating $\psi_{xx}\sim\frac{1}{c_0^2}\psi_{tt}$, one can derive the lossless Kuznetsov equation \cite{w},
\begin{equation}
\psi_{tt}-c_{s0}^2\psi_{xx}=2\psi_{xt}\psi_x+\frac{(\gamma-1)}{c_{s0}^2}\psi_{tt}\psi_t
\end{equation}
Studying nonlinear acoustic phenomena is not the aim of this paper. Hence we are not going into more details about it.\\
Thus the nonlinear wave in lossless regime can also be described as acoustic gravitational wave propagating in the medium. Here gravitational wave like effect is the emergent phenomena in the system.
\section{Stratified medium}
\subsection{Isothermal stratified medium}
Let's consider a isothermal medium of uniform temperature $T_0$, having a density verification along $z$ axis due to the external force along $z$ direction. For example, in a constant gravitational field, $g$, acting along $-z$ direction, the density $\rho(z)=\rho(0)exp\left(-\frac{z}{(RT_0/gM_A)}\right)$ \cite{n}. In the absence of any perturbation in such a system, we have
\begin{equation}
\frac{1}{\rho_0}\frac{dp_0}{dz}+F_{ext}(z)=0
\end{equation}
$F_{ext}(z)$ is the external body force. Let's consider perturbations in the medium along $x$ direction.
\begin{eqnarray}
&\rho(x,y,z,t)=\rho_0(z)+\rho'(x,y,z,t)\\
&v'^y=v'^z=0\\
&v'^x=v'^x(x,t)
\end{eqnarray}
Equation (43) is compatible with irrotationlity condition.
The system has a preferred direction, i.e. $z$ direction. As at the very outset, we are assuming perturbations propagating along $x$ direction, we expect to get $2\times2$ acoustic metric instead of $4\times4$.  
Therefore, we have continuity equation and Euler momentum equation for the perturbed quantities as follows
\begin{eqnarray}
\partial_t\rho'+\partial_x(\rho v'^x)=0\\
\partial_t v'^x+\partial_x\zeta'=0
\end{eqnarray}
The nonlinear perturbation $\partial_t\zeta'=v'^x\partial_tv'^x+\frac{c_s^2}{\rho}\partial_t\rho'$. After some manipulations we find
\begin{equation}
\partial_\mu(f^{\mu\nu}\partial_\nu)\zeta'=0
\end{equation}
where $\mu$, $\nu$ run over $t$ and $x$. $f^{\mu\nu}$ is given by
\begin{equation}
f^{\mu\nu}(x,z,t)\equiv\frac{\rho}{c_{s}^2}\begin{bmatrix}
-1 & -v'^x \\
-v'^{x}& c_{s}^{2}-(v'^{x})^2
\end{bmatrix} 
\end{equation}
Defining $(g_{\mu\nu})_{eff}$ \cite{x}, as the problem is not intrinsically $1+1$ dimensional rather it's $3+1$ dimensional \cite{y}; we can write \footnote{As the actual dimension (as in section 2) of the problem is $3+1$ because in this problem, the equation, $\partial_\mu(f^{\mu\nu}\partial_\nu)\zeta'=0$, still holds even for general perturbations, i.e. in the presence of nonzero $v'^x,~v'^y$ and $v'^z$; here in the very beginning choosing the symmetry and the direction of the wave reduces the dimension of the wave equation to $1+1$. Alternatively, for wave propagating in arbitrary direction, after deriving $\partial_\mu(f^{\mu\nu}\partial_\nu)\zeta'=0$, we could have chosen the symmetry and the direction of the wave  as we did in section 2. Therefore we use the same conformal factor in front of the metric as before.}
\begin{equation}
(g_{\mu\nu})_{eff}\equiv\frac{\rho}{c_s}\begin{bmatrix}
 -(c_{s}^{2}-v'^{2}) &-v'^{x} \\
-v'^{x} & 1
\end{bmatrix}
\end{equation}
After getting rid of the conformal factor, considering expansion of the terms upto second order and considering the disturbances to be of adiabatic type, we find in the similar fashion as before
\begin{eqnarray}
&\tilde{g}_{\mu\nu}({\bf{x}}, t)=(\eta_A)_{\mu\nu}+h_{\mu\nu}({\bf{x}}, t)\\
&\equiv\begin{bmatrix}
 -c_{s0}^{2}\left(1+\frac{(\gamma+1)}{2}\frac{\rho'_{(1)}}{\rho_0}\right)& -v'^{x}_{(1)} \\
-v'^{x}_{(1)} &\left(1+\frac{(3-\gamma)}{2}\frac{\rho'_{(1)}}{\rho_0}\right)
\end{bmatrix}
\end{eqnarray}
As the medium has uniform constant temperature in the absence of any disturbances, the linear sound speed, $c_{s0}(=\sqrt{\frac{\gamma RT_0}{M_A}})$ is same everywhere.\\
Considering, continuity equation and Euler equation upto first order of smallness,
\begin{eqnarray}
&\partial_t\rho'_{1}+\partial_x(\rho_0 v'^x_{1})=0\\
&\partial_tv'^x_{1}+\frac{c_{s0}^2}{\rho_0}\partial_x\rho'_1=0\\
&\partial_z(\frac{c_{s0}^2}{\rho_0}\rho'_1)=0
\end{eqnarray}
$\because~v'^x_{1}=v'^x_{1}(x,t)$ and $\rho_0=\rho_0(z)$, $\rho'_1(x,y,z,t)=\frac{\rho_0(z)}{c_{s0}^2}\epsilon(x,t)$ from the above equations. $\epsilon(x,t)$ is a function within the first order of smallness. Therefore
\begin{eqnarray}
&\partial_t^2\rho'_1=c_{s0}^2\partial_x^2\rho'_1\\
&\partial_t^2v'^x_1=c_{s0}^2\partial_x^2v'^x_1\\
&\Rightarrow \left(-\frac{1}{c_{s0}^2}\partial_t^2+\partial_x^2\right)h_{\mu\nu}(x,t)=0
\end{eqnarray}
The above equation has solution of plane wave propagating along $\pm x$ axis.\\
Here we have tacitly chosen the coordinate system first and we have assumed the perturbations across the perpendicular direction of stratification in the medium. We restrict ourselves by considering perturbations perpendicular to the direction of stratification. Nevertheless the form of equation (46) does not depend on the direction of perturbations but the form of equation (54)-(56) depends on the relative orientation between the direction of propagating wave and the direction of stratification. Unlike the previous case of uniform medium, in this case there is a preferred direction in the system, i.e. the direction of external body force, the symmetry is lost. That is why the wave propagating along the direction of stratification is different from the wave propagating across it. Even disturbances linear in nature propagating parallel to the direction of external body force have attenuation and would be of dispersive in nature \cite{n}. 
\subsection{Adiabatic stratified medium}
Let's consider the direction of external body force is along $z$ axis as before. In the absence of any perturbation, pressure $p_0(z)\propto\rho_0(z)^\gamma$. Unlike the previous case, the sound speed, more precisely the speed of linear perturbation, is not a constant number rather a function of $z$. For example, in the case of adiabatic medium in a constant gravitational field, $-g\hat{z}$; sound speed diminishes linearly with $z$ as the temperature diminishes linearly with height, $z$.   
\\
Introducing perturbations in the medium as below,
\begin{eqnarray}
&\rho(x,z,t)=\rho_0(z)+\rho'(x,z,t)\\
&v^x(x,z,t)=v'^x(x,z,t)\\
&v^z(x,z,t)=v'^z(x,z,t)\\
\end{eqnarray}
such that $\partial_xv'^z(x,z,t)=\partial_zv'^x(x,z,t)$, i.e. the irrotationality condition is satisfied \footnote{If one chooses simply as before, $v'^z=0$ and $v'^x=v'^x(x,t)$, in the same manner one could derive $\partial_t^2v'^x_1=c_{s0}^2\partial_x^2v'^x_1$ which does not make any sense because $c_{s0}=c_{s0}(z)$ whereas $v'^x=v'^x(x,t)$. That is why we choose $v'^x=v'^x(x,z,t)$ and to satisfy irrotationality condition we need $v'^z(x,z,t)$.} . Again after manipulation in similar fashion, one gets
\begin{equation}
\partial_\mu(f^{\mu\nu}(x,z,t)\partial_\nu)\zeta'(x,z,t)=0
\end{equation}
where
\begin{equation}
f^{\mu\nu}(x,z,t)\equiv\frac{\rho}{c_{s}^2}\begin{bmatrix}
-1 & -v'^x & -v'^z\\ 
-v'^{x}& c_{s}^{2}-(v'^{x})^2 & -v'^xv'^z\\
-v'^{z} & -v'^{z}v'^{x} & c_{s}^{2}-(v'^{z})^2
\end{bmatrix} 
\end{equation}
We get a three dimensional matrix, because we choose the quantities having dependence on one time dimension and two spatial dimension.
The continuity equation and Euler equation in the first order of smallness,
\begin{eqnarray}
&\partial_t\rho'_{1}+\partial_x(\rho_0v'^x_{1})+\partial_z(\rho_0v'^z_{1})=0\\
&\partial_tv'^x_{1}+\frac{c_{s0}^2}{\rho_0}\partial_x(\rho'_1)=0\\
&\partial_tv'^z_{1}+\partial_z(\frac{c_{s0}^2}{\rho_0}\rho'_1)=0
\end{eqnarray} 
From equation in the zeroth order of smallness, we have
\begin{equation}
\frac{1}{\rho_0(z)}\frac{dp_0(z)}{dz}=-F_{ext}(z)=\frac{d\Phi(z)}{dz}
\end{equation}
where $\Phi(z)$ corresponds to the potential corresponding to the conservative external body force. Hence at height $\frac{l}{2}$ from the height $z_0$, at
\begin{equation}
\rho_0(z_0+\frac{l}{2})=\rho_0(z_0)\left(1+ \frac{1}{c_{s0}^2(z_0)}\left(\Phi(z_0+\frac{l}{2})-\Phi(z_0)\right)+...\right)
\end{equation}
Free fall velocity from height $z_0+\frac{l}{2}$ to $z_0$ is $\sqrt{2\left(\Phi(z_0+\frac{l}{2})-\Phi(z_0)\right)}$. Hence $\rho_0(z_0+\frac{l}{2})\sim\rho_0(z_0)$ when the height difference is such that the free fall velocity is negligible compared to the sound speed within one's tolerance range of precision. Therefore, $\rho'_1(x,z_0+\frac{l}{2},t)\sim\rho'_1(x,z_0,t)$. We assume $v'^z(x,z,t)$ to be slowly varying function of $z$ which means $v'^z(x,z_0+\frac{l}{2},t)\sim v'^z(x,z_0,t)$. Basically, we are trying to generate wave propagating perpendicular to the direction of stratification which follows $v'^x(x,z_0,t)\sim v'^x(x,z_0+\frac{l}{2},t)$, otherwise the viscous effects due to shear force, would come into play. From irrotationality condition, $v'^z(x,z,t)$ is very small. As the sound speed is also more or less same within the slice of space between heights, $z_0+\frac{l}{2}$ and $z_0-\frac{l}{2}$, we effectively reduce the problem to a problem of isothermal medium. Two things are done here simultaneously, one is that disturbance propagating perpendicular to the direction of stratification is chosen and secondly a slice of space having a thickness along the direction of external body force is chosen in such a way that all the variations along $z$ becomes negligible.\\
Averaging out equations (63) to (65) by integrating over $z$, over $z$ in the limit $z_0-\frac{l}{2}$ to $z_0+\frac{l}{2}$, writing average of $\rho'_1(x,z,t)$ as $\tilde{\rho}'_1(x,t)\sim\rho'_1(x,z_0,t)$ and average of $v'^x_{1}(x,z,t)$ as $\tilde{v'_x}(x,t)\sim v'^x(x,z_0,t)$, we get
\begin{eqnarray}
&\partial_t\tilde{\rho}'_{1}+\partial_x(\rho_0(z_0) \tilde{v}'^x_{1})=0\nonumber\\
&\partial_t\tilde{v}'^x_{1}+\frac{c_{s0}^2(z_0)}{\rho_0(z_0)}\partial_x\tilde{\rho}'_1=0\nonumber\\
&\Rightarrow\partial_t^2\tilde{\rho}'_1=c_{s0}^2(z_0)\partial_x^2\tilde{\rho}'_1\\
&\Rightarrow\partial_t^2\tilde{v}'^x_1=c_{s0}^2(z_0)\partial_x^2\tilde{v}'^x_1
\end{eqnarray}
Similarly averaging $f^{\mu\nu}$ over $z$ and rewriting it as $\tilde{f}^{\mu\nu}$;
\begin{equation}
\tilde{f}^{\mu\nu}(x,t)\equiv\frac{\tilde{\rho}}{\tilde{c_{s}}^2}\begin{bmatrix}
-1 & -\tilde{v}'^x & 0\\ 
-\tilde{v}'^{x}& \tilde{c_{s}}^{2}-(\tilde{v}'^{x})^2 & 0\\
0 & 0 & \tilde{c_{s}}^{2}-(\tilde{v}'^{z})^2
\end{bmatrix} 
\end{equation}
Now finding the acoustic metric after getting rid of the conformal factor and carrying the expressions upto second order of smallness, we find using equation (68) and equation (69) in the same manner as done before
\begin{equation}
\left(-\frac{1}{c_{s0}^2(z_0)}\partial_t^2+\partial_x^2\right)h_{\mu\nu}(x,t)=0
\end{equation} 
\section{Shallow water wave}
Shallow water wave or long gravity wave, i.e. the wavelength of such wave is longer than the depth of the incompressible liquid medium \cite{z}. We assume constant gravity, $-g\hat{z}$; depth of the liquid is denoted by $h$. We consider shallow water wave, not necessarily linear, the liquid flows through a channel (along $x$ axis) and the wave is longitudinal \cite{z}, i.e. the velocity along $z$ direction and $y$ direction is much smaller compared to the velocity, $v$ along $x$ axis; the continuity equation reads as \cite{z}
\begin{equation}
\frac{\partial h}{\partial t}+\frac{\partial}{\partial x}(vh)=0
\end{equation} 
The Euler momentum equation has the form
\begin{equation}
\frac{\partial v}{\partial t}+v\frac{\partial v}{\partial x}+g\frac{\partial h}{\partial x}=0
\end{equation}
Now we define a quantity $\xi$ as $(\frac{1}{2}v^2+gh)$ which is very similar to Bernoulli's constant $\zeta$ in the previous cases. Hence from the momentum equation,
\begin{equation}
\frac{\partial v}{\partial t}+\frac{\partial \xi}{\partial x}=0
\end{equation}
The fluid velocity and height are given by
\begin{eqnarray}
&v(x,t)=v'(x,t)\\
&h(x,t)=h_0+h'(x,t)\\
\end{eqnarray}
where primed quantities are the perturbations in the system. $h_0$ is the constant height of the liquid in the absence of any disturbances. The sound speed corresponding to linear perturbation, $c_{s0}$ is $\sqrt{gh_0}$ \cite{z}. After manipulations with the perturbed quantities in the same manner as before; we find
\begin{equation}
\partial_\mu(f^{\mu\nu}\partial_\nu(x,t))\xi'=0
\end{equation}
where $\mu$, $\nu$ run over $t$ and $x$. $f^{\mu\nu}$ is given by
\begin{equation}
f^{\mu\nu}(x,t)\equiv\begin{bmatrix}
-1 & -v' \\
-v'& gh-v'^2
\end{bmatrix} 
\end{equation}
$gh$ in the matrix can be denoted as $c_s^2$.
There is no conformal factor in front of $f^{\mu\nu}$ because in this problem $h$ is mathematically equivalent to $\rho$ in the previous problems and $c_s^2$ is proportional to $h$; that is why the conformal factor $\frac{\rho}{c_s^2}$ in the previous cases happen to be a constant number in this problem. Hence effectively, the acoustic metric becomes
\begin{equation}
g_{\mu\nu}(x,t)\equiv\sqrt{h}\begin{bmatrix}
 -(c_{s}^{2}-v'^{2}) &-v'^{x} \\
-v'^{x} & 1
\end{bmatrix}
\end{equation}
The conformal factor is $\sqrt{h}$ because in the previous cases, the conformal factor $\frac{\rho}{c_s}$ in front of $g_{\mu\nu}$ is equivalent to $\frac{h}{\sqrt{gh}}$ here. 
After expanding the equations upto second order of smallness, i.e, in the weak nonlinear limit; we find after dropping the conformal factor $\sqrt{h_0}$;
\begin{equation}
\tilde{g}_{\mu\nu}(x,t)=(\eta_A)_{\mu\nu}+h_{\mu\nu}(x,t)\equiv\begin{bmatrix}
 -c_{s0}^2\left(1+\frac{3}{2}\frac{h'_1}{h_0}\right) &-v'^{x}_1 \\
-v'^{x}_1 & \left(1+\frac{1}{2}\frac{h'_1}{h_0}\right)
\end{bmatrix}
\end{equation}
One can show in the same manner that
\begin{equation}
\left(-\frac{1}{c_{s0}^2}\partial_t^2+\partial_x^2\right)h_{\mu\nu}(x,t)=0
\end{equation}
Shallow water wave in weakly nonlinear limit can also be experimentally realized \cite{a1}. 
\section{Bose Einstein Condensate in a tight ring trap}
A dilute very cold (temperature $\sim 0 K$) weakly interacting BEC can be described by a classical field, $\Phi({\bf{x}},t)$, having the meaning of the order parameter, satisfying the time dependent Gross-Pitaevskii equation \cite{a2};
\begin{equation}
i\hbar\frac{\partial \Phi({\bf{x}},t)}{\partial t}=\left(-\frac{\hbar^2}{2m}\nabla^2+V_{ext}({\bf{x}})+g|\Phi({\bf{x}},t)|^2\right)\Phi({\bf{x}},t)
\end{equation}
where $V({\bf{x}})$ is the external potential, $g$ is two body interaction coefficient related to s-wave scattering cross-section.
\begin{align*}
g=\frac{4\pi\hbar^2a}{m}
\end{align*}
where $a$ is the scattering length. $g$ is positive for repulsive interaction and negative for attractive interaction. 
Stationary state, $\Phi_s({\bf{x}},t)$, satisfying eigenvalue equation, i.e. the time independent Gross-Pitaevskii equation; given by 
\begin{equation}
\left(-\frac{\hbar^2}{2m}\nabla^2+V_{ext}({\bf{x}})+g|\Phi_s({\bf{x}},t)|^2\right)\Phi_s({\bf{x}},t)=\mu \Phi_s({\bf{x}},t)
\end{equation}
where $\mu$ is the eigenvalue of the problem which is also the chemical potential of the problem.\\
Hence from equation (83), $\Phi_s({\bf{x}},t)=\Phi_s({\bf{x}},0)e^{-i\frac{\mu t}{\hbar}}=\sqrt{n_0({\bf{x}})}e^{iS({\bf{x}})}e^{-i\frac{\mu t}{\hbar}}$ with $i=\sqrt{-1}$, $n_0({\bf{x}})$ being the condensate number density, and $S({\bf{x}})$ being a phase factor.  Super fluid speed (resistance less speed) of BEC is proportional to the gradient of $S({\bf{x}})$ \cite{a2}. The energy functional $E[\Phi]$ is given by \cite{a2}
\begin{equation}
E[\Phi]=\bigintsss d^3{\bf{x}}\left(\frac{\hbar^2}{2m}\mid\nabla\Phi\mid^2+V_{ext}({\bf{x}})\mid\Phi\mid^2+\frac{g}{2}\mid\Phi\mid^4\right)
\end{equation} 
The first term, second term and the third term in the integral correspond to the kinetic energy ($E_{kin}$), the potential energy ($E_V$) and the interaction energy ($E_{int}$) respectively.\\
Ring traps as external potential are experimentally realized in many cases \cite{a3}-\cite{a7}. Here we discuss about toroidal ring trap given by 
\begin{equation}
V_{ext}({\bf{x}})=\frac{1}{2}m\omega^2\left((r-R)^2+z^2\right)
\end{equation}
For simplicity, we assume the trapping frequency  along the cylindrical radial direction $r$ is same as the trapping frequency along $z$, denoted by $\omega$. 
\subsection{The energy scales and the length scales of the problem}
In the ground state, i.e. the state with zero superfluid speed; $S({\bf{x}})$ can be assumed to be zero. Hence the  solution of stationary GP equation is effectively a function of density, $n_0$ only, i.e.  $\Phi_s({\bf{x}},t)=\sqrt{n_0({\bf{x}})}e^{-i\frac{\mu t}{\hbar}}$. Therefore the energy is a functional of number density only \cite{a2}.
\begin{equation}
E[\sqrt{n_0}]=\bigintsss d^3{\bf{x}}\left(\frac{\hbar^2}{2m}\mid\nabla\sqrt{n_0}\mid^2+V_{ext}({\bf{x}})n_0+\frac{g}{2}n_0^2\right)
\end{equation}
Length scale along the $z$ direction and $r$ direction around the radius $R$, i.e around the circle of minima of the potential on $z=0$ plane, is $a_{ho}=\left(\frac{\hbar}{m\omega}\right)^{1/2}$ \cite{a2}. Length scale along the azimuthal direction is $R$. Hence the volume scale of the problem is $a_{ho}^2R$, $N\sim\bar{n}_0a_{ho}^2R$ where $N$ is the total number of atoms in the trap. Therefore, we have
\begin{eqnarray}
& \mid\Phi_s\mid\sim\sqrt{\bar{n}_0},~\mid\frac{\partial\Phi_s}{\partial r}\mid\sim\mid\frac{\partial\Phi_s}{\partial z}\mid\sim\frac{\sqrt{\bar{n}_0}}{a_{ho}},~\mid\frac{\partial\Phi_s}{r\partial \varphi}\mid\sim\frac{\sqrt{\bar{n}_0}}{R}\nonumber
\end{eqnarray}
where $\bar{n}_0$ is the spatially average number density of condensate atoms.
Therefore,
\begin{equation}
E_{int}\sim g\bar{n}_0N=\frac{4\pi\hbar^2a}{m}\bar{n}_0N,~E_{kin}^{\varphi}\sim\left(\frac{\hbar^2}{2m}\right)\frac{\bar{n}_0a_{ho}^2}{R},~E_{kin}^{r}\sim E_{kin}^z\sim \frac{\hbar^2}{2m}\bar{n}_0R \nonumber
\end{equation}
where $E_{kin}^{\varphi},~E_{kin}^{r}$ and $E_{kin}^z$ are the kinetic energy components along $\varphi$, $r$ and $z$ directions respectively.
\begin{eqnarray}
&\Rightarrow \frac{E_{kin}^{r,z}}{E_{int}}\sim\frac{R}{Na}\\
&\Rightarrow\frac{E_{kin}^{\varphi}}{E_{int}}\sim\left(\frac{a_{ho}}{R}\right)\left(\frac{a_{ho}}{Na}\right)
\end{eqnarray} 
For a tight toroidal trap, $a_{ho}\ll R$. 
\subsection{Ground state solution of the stationary GP equation}
We seek solution for the ground state, as the external potential does not depend on the azimuthal angle $\varphi$, therefore $\Phi_s({\bf{x}},t)=\Phi_s(r,z,t)$. Under T-F approximation, the GP equation can be reduced to classical fluid equations \cite{a2}. We assume the bosonic atoms to be strongly repulsively interacting, i.e. the T-F approximation, $E_{int}>>E_{kin}$. As there is no azimuthal angle dependence of the ground state function, from the above section, $T-F$ approximation in this case means $Na\gg R$ which automatically implies for a tight toroidal trap $Na\gg R\gg a_{ho}$. Therefore, dropping the kinetic term in the equation (84), we find
\begin{equation}
n_0(r,z)=\frac{\mu-V_{ext}(r,z)}{g}
\end{equation}
$n_0(r,z)>0$ for $\mu> V_{ext}(r,z)$ and zero for $\mu\leq V_{ext}(r,z)$.\\
We define a new coordinate system as
\begin{eqnarray}
& (r-R)=\chi cos\alpha\\
& z=\chi sin\alpha
\end{eqnarray}
where $\chi$ is the distance from $r=R$ at a fixed $\phi$; $\alpha$ is the angle of that distance with $r-\varphi$ plane. 
Therefore $V_{ext}(r,z)=V_{ext}(\chi)=\frac{1}{2}m\omega^2\chi^2$. $n_0(\chi)~\left(=\frac{\mu-\frac{1}{2}m\omega^2\chi^2}{g}\right)$ is greater than zero for $\chi<\chi_0(=\frac{\sqrt{2\mu/m}}{\omega})$ and is zero for $\chi\geq\chi_0$. Therefore under T-F approximation, the BEC atoms are confined within a torus of finite radius $\chi_0$ surrounding the minima of the potential function at $r=R$ on $z=0$ plane.  $\chi_0$ is determined by the equation
\begin{equation}
N=\int_V n_0 d^3{\bf{x}}=2\pi\int_0^{2\pi}\int_0^{\chi_0} d\chi d\alpha\chi(R+\chi cos\alpha)\frac{1}{g} \left(\mu-\frac{1}{2}m\omega^2\chi^2\right)
\end{equation}
$V$  is the volume of the torus with $\chi_0$ being the radius of it's circular section.
\begin{figure}
\centering
\includegraphics[scale=0.35]{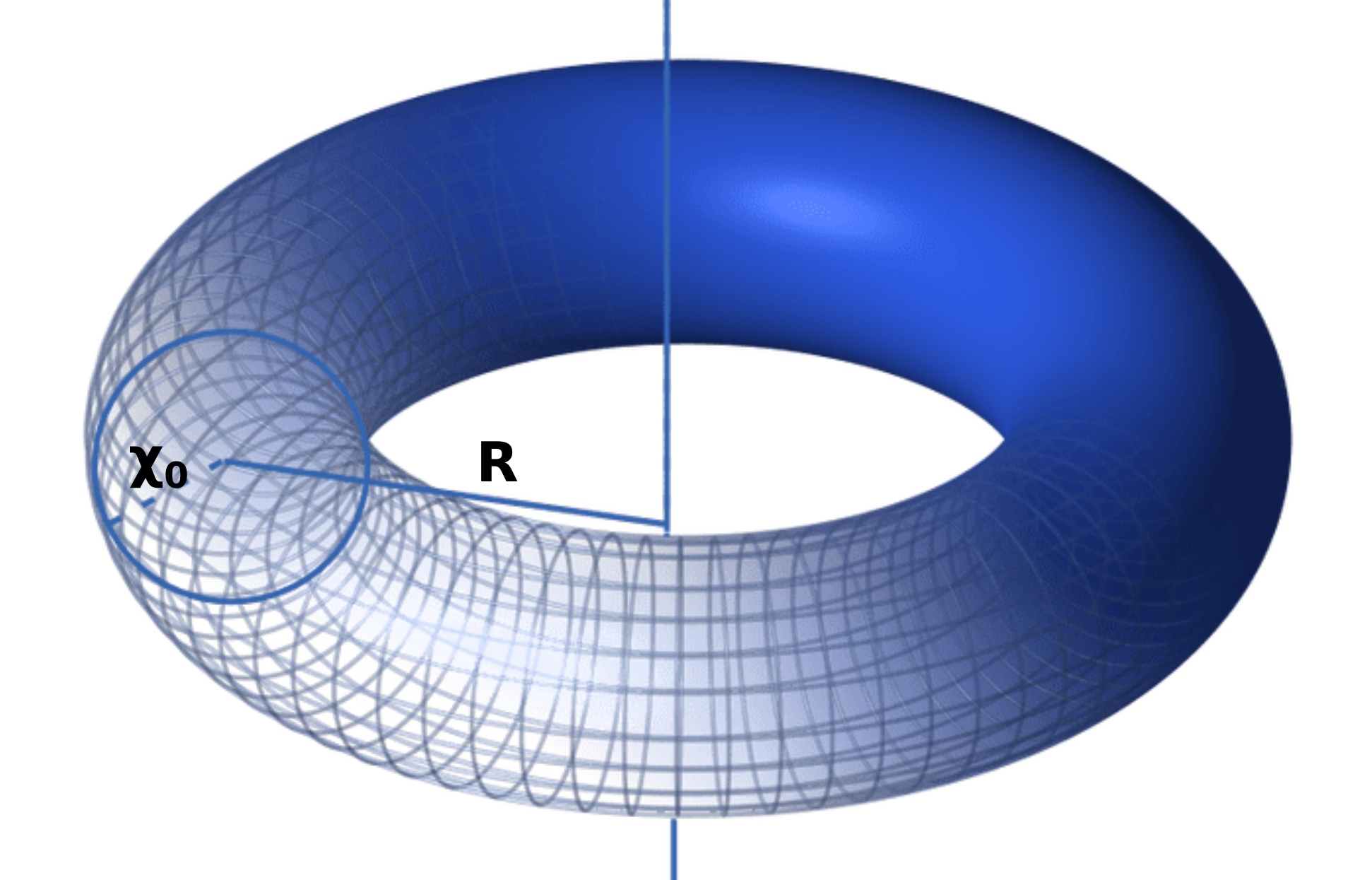}\includegraphics[scale=0.3]{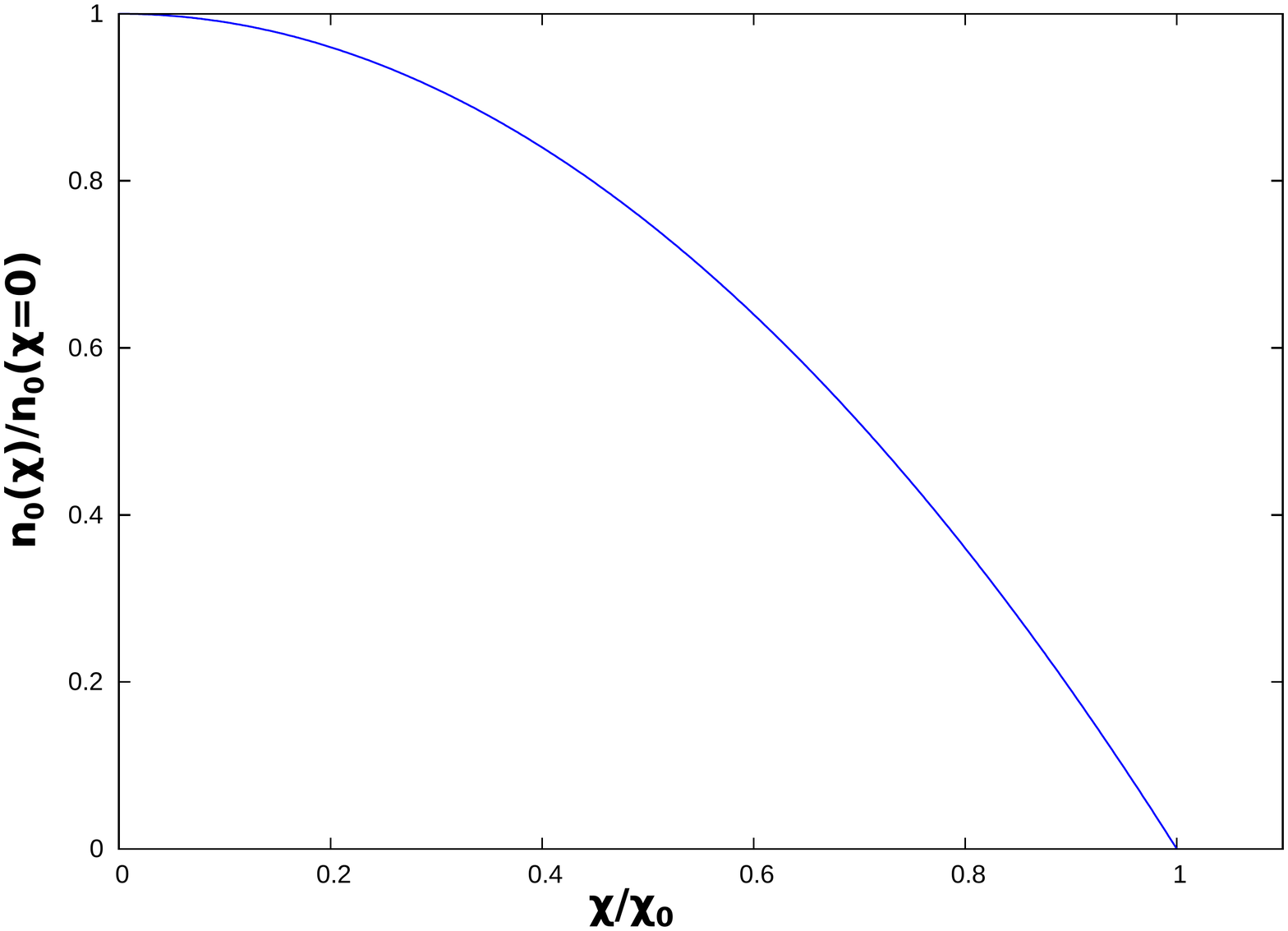}
\caption{Demonstrating the ring torus region of space (the blue region in the figure) within which all the BEC atoms are effectively trapped (left) and the Thomas-Fermi number density distribution (right).}
\end{figure}
$\mu~(=\frac{1}{2}m\omega^2\chi_0^2)$ determines the maximum $\chi$ radius within which all the atoms stay.
We find $\chi_0=2a_{ho}\left(\frac{aN}{2\pi R}\right)^{\frac{1}{4}}$.\\
Therefore
\begin{equation}
n_0(r,z)=n_0(\chi)=\frac{1}{8\pi a_{ho}^4a}\left(4a_{ho}^2\left(\frac{aN}{2\pi R}\right)^{\frac{1}{2}}-\chi^2\right)=\frac{1}{8\pi a_{ho}^4a}\left(4a_{ho}^2\left(\frac{aN}{2\pi R}\right)^{\frac{1}{2}}-(r-R)^2-z^2\right)
\end{equation}  
Therefore, from equation (9), stationary ground state solution of GP equation is
\begin{equation}
\Phi_s(r,z,t)=\sqrt{n_0(r,z)}e^{-i\frac{\mu t}{\hbar}}
\end{equation}
\subsection{Perturbative approach to the time dependent GP equation}
With tight radial and axial component, the dynamics along the radial direction and the axial direction is 'frozen' \cite{a8}. The problem becomes effectively one dimensional. Therefore, the wave function can be written as
\begin{equation}
\Phi=f(r,z)\psi(\varphi,t)
\end{equation}
where $\int_B rdrdz f(r,z)^2=1$ and $\int_0^{2\pi}d\varphi\mid\psi\mid^2=N$, where $B=\frac{V}{2\pi}=\pi\chi_0^2R$.
\begin{eqnarray}
 & i\hbar f(r,z)\frac{\partial \psi(\varphi,t)}{\partial t}=-\frac{\hbar^2}{2m}\left(\frac{\psi}{r}\frac{\partial }{\partial r}\left(r\frac{\partial}{\partial r}\right)f(r,z)+\frac{f}{r^2}\frac{\partial^2}{\partial\varphi^2}\psi(\varphi,t)+\psi\frac{\partial^2}{\partial z^2}f(r,z)\right)  \nonumber\\
 & +\frac{1}{2}m\omega^2\left((r-R)^2+z^2\right)f\psi+g|f|^2|\psi|^2f\psi
\end{eqnarray}
Now we insert the expression of stationary ground state solution in the above equation as below
\begin{equation}
f(r,z)=\sqrt{\frac{2\pi}{N}}\sqrt{n_0(r,z)}
\end{equation}
Therefore, due to T-F approximation, the first and the third term in the right hand side of the equation (97) vanishes. Hence 
\begin{equation}
  i\hbar f\frac{\partial \psi(\varphi,t)}{\partial t}=-\frac{\hbar^2}{2m}\frac{f}{r^2}\frac{\partial^2}{\partial\varphi^2}\psi(\varphi,t) 
  +\left(\frac{1}{2}m\omega^2\left((r-R)^2+z^2\right)+g|f|^2|\psi|^2\right)f\psi
\end{equation}
The dependence on $r$ and $z$ can be projected out by multiplying the above equation by $f^*$ and integrating the above equation over $r$ and $z$ within the volume $B$. We find
\begin{equation}
i\hbar \frac{\partial \psi}{\partial t}=-\left(\frac{\hbar^2}{2mR^2}\right)\frac{\partial^2\psi}{\partial\varphi^2} 
  +\left(\frac{\mathfrak{n}}{12\bar{n}_0}\right)m\omega^2\chi_0^2\psi+\tilde{g}|\psi|^2\psi
\end{equation}
where $\tilde{g}=g\frac{2\pi \mathfrak{n}^2}{3N\bar{n}_0}$, $\mathfrak{n}=n_0(\chi=0)$ and the average number density, $\bar{n}_0=\frac{N}{2\pi R\pi\chi_0^2}$. The factor $\frac{1}{R^2}$ in the first term of the right hand side is appearing because we are considering the wave function to be concentrated around the minima circle of potential due to tightness of the trap \cite{a8}. Thus the problem becomes effectively one dimensional. The second term in the right hand side of the equation is a constant shift in potential, we make it zero by translation in the potential. Therefore, finally we have
\begin{equation}
i\hbar \frac{\partial \psi}{\partial t}=-\frac{\hbar^2}{2mR^2}\frac{\partial^2\psi}{\partial\varphi^2} 
  +\tilde{g}|\psi|^2\psi
\end{equation} 
We decompose $\psi$ as
\begin{equation}
\psi=\psi_S+\psi'(\varphi,t)
\end{equation}
$\psi_S$ corresponds to $\psi$ in the ground state which is proportional to $e^{-i\frac{\mu t}{\hbar}}$. The number density, $n(\varphi,t)=|\psi|^2={\rm const}+n'(\varphi,t)$ and $\psi=\sqrt{n}e^{i\gamma(\varphi,t)}$; where $n'(\varphi,t)$ is the perturbation in number density which is not necessarily linear. The velocity, along $\varphi$, $v_\varphi$ is proportional to $\frac{\partial \gamma}{\partial\varphi}$. Putting this value of $\psi$ and using $T-F$ approximation, we get classical inviscid irrotational fluid equations \cite{a2}
\begin{eqnarray}
& \partial_t \rho+\frac{1}{R}\frac{\partial}{\partial\varphi}(\rho v)=0\\
& {\bf{\nabla}}\times{\bf{v}}=0\\
& \frac{\partial v_\varphi}{\partial t}+\frac{v_\varphi}{R}\frac{\partial v_\varphi}{\partial \varphi}=-\frac{1}{\rho R}\frac{\partial p}{\partial\varphi}\\
\end{eqnarray}
where $p=\frac{1}{2}\tilde{g}n^2$, sound speed, $c_{s0}=\sqrt{\frac{\tilde{g}n}{m}}$.  
We have
\begin{equation}
\rho={\rm const}+\rho'
\end{equation}
and
\begin{equation}
v_\varphi=v_\varphi'
\end{equation}
After manipulations in same manner as before, we find
\begin{equation}
\partial_\mu(f^{\mu\nu}(\varphi,t)\partial_\nu)\zeta'(\varphi,t)=0
\end{equation}
The Greek indices in the above equation run over time $t$ and the compact dimension, $\mathscr{R}=R\varphi$. 
\begin{equation}
f^{\mu\nu}(\varphi, t)\equiv\frac{\rho}{c_{s}^2}\begin{bmatrix}
-1 & -v'_\varphi\\
-v'_\varphi & c_{s}^{2}-(v'_\varphi)^2
\end{bmatrix} 
\end{equation}
and
\begin{equation}
(g_{\mu\nu})_{eff}\equiv\frac{\rho}{c_s}\begin{bmatrix}
 -(c_{s}^{2}-(v'_\varphi)^{2}) &-v'_\varphi \\
-v'_\varphi & 1
\end{bmatrix}
\end{equation}
Proceeding in the same fashion, we find
\begin{equation}
\tilde{g}_{\mu\nu}(\mathscr{R},t)=(\eta_A)_{\mu\nu}+h_{\mu\nu}(\mathscr{R},t)\equiv\begin{bmatrix}
 -c_{s0}^2\left(1+\frac{3}{2}\frac{\rho'_1}{\rho_0}\right) &-(v'_\varphi)_1 \\
-(v'_\varphi)_1 & \left(1+\frac{1}{2}\frac{\rho'_1}{\rho_0}\right)
\end{bmatrix}
\end{equation}
This is very similar to equation (81). 
One can show in the same manner that
\begin{equation}
\left(-\frac{1}{c_{s0}^2}\partial_t^2+\partial_\mathscr{R}^2\right)h_{\mu\nu}(\mathscr{R},t)=0
\end{equation}
Here the difference from the previous cases is that the spatial dimension is a compact dimension.  Any perturbation produced in the toroidial ring will propagate in clockwise and anticlockwise senses and eventually will superimpose with each other and thus standing wave will be produced. Here we can view the scenario as the standing wave of acoustic analogue gravitational wave.\\
One can do same kind of analysis in other tight traps of different geometries. The methods would be very similar to the method discussed in this section. 
\section{Summary and conclusions}
We find that if one extends the perturbative method of analysis in the inviscid irrotational fluid equations of finding the acoustic space-time geometry in weakly nonlinear limit for static fluid systems as background, one discovers that the space-time geometry of the acoustic metric happens to get some properties which are very similar to the acoustic metric describing gravitational wave propagation in Minkowski spacetime. The transverse nature of grvitational wave is missing in the acoustic metric; rather it describes longitudinal wave in the acoustic analogue of Minkowski spacetime. Our analysis also makes a connection between two seemingly different subjects; one is nonlinear acoustics and the other one is the study of emergent spacetime. In weakly nonlinear limit, the acoustic analogue of gravitational wave is the emergent phenomena. 

\newpage
\section*{Appendix}
\subsection*{Perturbations in mass flow rate}
Mass flow rate is defined as the amount of mass of fluid passing through a unit area perpendicularly per unit time and that is why we first specify the direction of the disturbance and we consider the medium to be uniform in the absence of any perturbations. 
\begin{align*}
& p(z, t)=p_0+p'(z, t)\\
& \rho(z, t)=\rho_0+\rho'(z, t)\\
& v^z=v'^z(z, t)\\
& v^x=0\\
& v^y=0
\end{align*}
Therefore, mass flow rate, $f=\rho v^z=f'(z,t)=\rho v'^z$. From the continuity equation and the Euler equation, can be written as
\begin{equation}
\frac{\partial\rho'}{\partial t}+\partial_z(f')=0
\end{equation}
\begin{equation}
\partial_{tt} v'^z+\partial_z(v'^z\partial_t v'^z)=\partial_z\left(\frac{c_s^2}{\rho}\partial_t f'\right)
\end{equation}
From the definition of the mass accretion rate
\begin{equation}
\partial_t v'^z=\frac{1}{\rho}\left(\partial_t f'+v'^z\partial_z f'\right)
\end{equation}
Therefore, using equation (115) and equation (116), we get
\begin{equation}
\partial_\mu (f^{\mu\nu}(z,t)\partial_\nu )f'(z,t)=0
\end{equation}
where 
\begin{equation}
f^{\mu\nu}(z,t)\equiv\frac{1}{\rho}\begin{bmatrix}
-1 &  -v'^z \\
-v'^{z}& c_{s}^{2}\delta^{ij}-(v'^z)^2
\end{bmatrix}
\end{equation}
The above matrix is a $2\times 2$ matrix because we have chosen the direction of the perturbation first. One can not derive $g^{\mu\nu}$ from $f^{\mu\nu}=\sqrt{-g}g^{\mu\nu}$ for $2\times 2$ matrices. As the problem is intrinsically $3+1$ dimensional, we conventionally use the same $g^{\mu\nu}$ obtained from the wave equation of Bernoulli's constant. As a result, the rest of the analysis becomes same.
\end{document}